# $Au^{9+}$ swift heavy ion irradiation of $Zn[CS(NH_2)_2]_3SO_4$ crystal: Crystalline perfection and optical properties

S. K. Kushwaha[1,2], K. K. Maurya[1], N. Vijayan[1], A. K. Gupta[1], D. Haranath[1], B. Kumar[3], D. Kanjilal[4] and G. Bhagavannarayana[1]

[1]CSIR - National Physical Laboratory, Dr. K. S. Krishnan Road, New Delhi, 110012, India
[2]Department of Chemistry, Princeton University, Princeton NJ 08544, USA
[3]Department of Physics & Astrophysics, University of Delhi, Delhi, 110007, India
[4]Inter-University Accelerator Centre, New Delhi, 110067, India

The single crystal of tris(thiourea)zinc sulphate ($Zn[CS(NH_2)_2]_3SO_4$) was irradiated by 150 MeV $Au^{9+}$ swift heavy ions and analyzed in comparison with pure crystal for crystalline perfection and optical properties. The Fourier transform infrared and x-ray powder diffraction inferred that swift ions lead the disordering and breaking of molecular bonds in lattice without formation of new structural phases. High resolution X-ray diffraction (HRXRD) revealed the abundance of point defects, and formation of mosaics and low angle grain boundaries in the irradiated region of crystal. The swift ion irradiation found to affect the lattice vibrational modes and functional groups significantly. The defects induced by heavy ions act as the color centers and resulted in enhance of photoluminescence emission intensity. The optical transparency and band gap found to be decreased.

## 1. INTRODUCTION

The nonlinear optical (NLO) single crystals play an important role in the area of photonics for communication, high density data storage, retrieval, processing and transmission of data and laser technology [1,2]. The design of photonic devices opened demand to search new NLO materials, to grow their single crystals and to improve or modify their structural and optical properties by suitable means for desired tailor made applications [3]. The swift heavy ions (SHI) irradiation leads to modification of the materials through the transfer of considerable amount of energy via collision with the electrons of target which result in the formation of high energy densities along the ion path [4-6]. The SHI lose their energy to the targeting medium through two main processes; (i) nuclear energy loss $(dE/dx)_n$ which is an elastic collision process and dominates for ions of energy ~1 keV/amu, and (ii) electronic energy loss $(dE/dx)_e$ which is an inelastic process and dominates for energy ~1MeV/amu. In the later process the energy transfer from SHI to the atoms of target medium takes place through the excitation and ionization, and in each collision the electronic energy loss ranges from tens of eV to a few keV per angstrom. This huge energy transfer leaves the target medium with specific effect [7-9], including the in-situ heating leading to recrystallization of amorphous surface layer by the process of solid state epitaxy [10]. The nuclear process results in the displacement of the atoms from their lattice position in the medium (Lesueur & Dunlop, 1993). The SHI significantly modify the optical properties of crystals by changing the refractive index [11]. Heavy ions are capable to break the molecular bonds and create damage to the single crystals [12]. SHI induced point defects in crystals act as the color centers [6] and lead to the enhancement in luminescence efficiency[13-14]. The crystallinity of crystals get modified and amorphization of materials also has been reported [15-16]. The structural defects play a vital role in defining the physical properties of crystal [17,18], therefore, in the modern technology of continuous miniaturization of devices, the crystalline perfection assessment is crucial.

However, no much literature is available about the investigation of crystalline perfection and optical properties of SHI irradiated NLO single crystals. Therefore, we aimed to investigate the SHI induced structural defects in tris(thiourea)zinc sulphate ($Zn[CS(NH_2)_2]_3SO_4$: ZTS) single crystal in correlation with its optical properties. The ZTS is well characterized promising NLO material which exhibits low angular sensitivity [19-20] and is suitable for the electro-optic applications [21]. In our recent studies we have observed an interesting correlation between the crystalline perfection and second harmonic generation efficiency (SHG) of ZTS crystals with different dopants [3,22,23].

In the present investigation, we report the influence of 150 MeV $Au^{9+}$ SHI on the crystalline perfection and optical properties of ZTS single crystal. The crystal structure and molecular bonds were studied by powder X-ray diffraction and Fourier transform infrared (FTIR) spectroscopy. The crystalline perfection was investigated by high resolution X-ray diffraction (HRXRD) employing a multicrystal X-ray diffractometer. The lattice vibrational modes and functional groups were analyzed by Raman spectroscopy and photoluminescence (PL) spectroscopy. The optical transparency, absorbance and band gap were studied and measured in comparison to the pristine crystal.

## 2. Experimental

The ZTS was synthesized in aqueous medium by taking thiourea and zinc sulphate in 3:1 molar ratio and the single crystals of ZTS were grown by solvent evaporation solution technique (SEST). The two identical crystals of dimensions $1\times3\times5$ mm$^3$ with natural facets were chosen for studies, one of them was subjected to SHI irradiation for the face corresponding to (200) diffraction planes. The irradiation experiment was performed on a 15-UD Pelletron at Inter University Accelerator Center (IUAC), New Delhi, India [24,25]. The crystal was mounted on a copper ladder with proper thermal contact under high vacuum of $10^{-6}$ Torr at ambient temperature. The SHI beam ($Au^{9+}$) was subjected to incident normally on the crystal face and the entire surface of the crystal was irradiated by scanning the Au ion beam. The energy, fluence and current of irradiating beam were maintained respectively at 150 MeV, $2\times10^{12}$ ions/cm$^2$ and 2 pnA throughout the experiment.

The functional groups of ZTS were analyzed using FTIR spectrometer (Nicolet-5700) in reflectance mode in the range of 400-3600 cm$^{-1}$ wavenumber and the the structure was confirmed by a Bruker AXS D8 Advanced Powder X-ray diffractometer with CuK$_\alpha$ radiation and having graphite monochromator. The diffraction patterns of crystals were recorded in identical experimental conditions in the 2-theta angular range of 10–80 degrees, with 0.01 °/s scan rate.

The detailed analysis of crystalline perfection of irradiated crystal in comparison to that of pristine was performed by a multicrystal X-ray diffractometer (MCD) [26] developed at National Physical Laboratory (NPL), India. In this system, a well-collimated and monochromated MoK$\alpha_1$ beam ($\Delta\lambda/\lambda\approx10^{-5}$) obtained from the Si monochromator crystals set in dispersive (+,-,-) configuration was used as an exploring X-ray beam [22,26]. The specimen crystal is aligned in the (+,-,-,+) configuration. The specimen was rotated about a vertical axis with a minimum angular step interval of 0.4 arc s. The rocking curves (RCs) for (200) diffraction planes of crystals were recorded by varying the glancing angle (angle between the incident X-ray beam and the surface of the specimen) around the Bragg diffraction peak ($\theta_B$), keeping the detector stationary at the angular position $2\theta_B$ with wide opening for its slit, the so called $\omega$ scan, whose details and advantages were explained in our recent report [22]. The results were compared with the theoretical limit of crystalline perfection of (200) diffraction planes of ZTS.

The lattice vibrational modes and functional groups were analysed by a Renishaw inVia Raman Microscope in back-scattering mode. A 785 nm wavelength diode laser (HP-NIR) was used as source with beam power of 300 mW, the spot size of laser beam on sample surface was 1 $\mu$m and the exposure time was 10 s. The beam was made to incident normally on the surface of (200) planes.

The photoluminescence behaviour was analysed by a Perkin Elmer LS-55, photoluminescence spectrophotometer. The optical transmittance, absorption coefficient and band gap were studied in the entire ultraviolet-visible-near infrared (UV-Vis-NIR) range (200-1100 nm) of energy spectrum, using a Perkin Elmer Lambda 35 spectrophotometer

## 3. Results and discussion

### 3.1. FTIR analysis

The SHI modify crystal lattice along their path by inducing heavy stresses and strains which results in the breaking, stretching and bending of molecular and intermolecular bonds [27]. In the modified regions of ion tracks, the molecules also undergo the relative disordering which cause the broader distribution of bond lengths and energies as compared to that of pristine crystal [28]. The recorded FTIR spectra of pristine as well as irradiated crystals are shown in Fig. 1 which consist of all the absorption peaks corresponding to the functional groups.

The bonds 1628, 1502, 1404 and 714 cm$^{-1}$ indicate NH$_2$ bending, N–C–N stretching, C=S asymmetric stretching and C=S symmetric stretching bonds, which are expected in ZTS crystals [22,29-31]. However, the intensity of all the absorption peaks reduced significantly and absorption bands get broadened for the irradiated crystal, which depicts the breaking and disordering of molecular bonds due to transfer of high energy to the crystal lattice from SHI. For clarity the magnified view of the spectra is shown in the inset of Fig.1, and the major bands have been assigned their corresponding vibrations and functional groups. The broadening of the absorption bands at 1400 cm$^{-1}$ and 700 cm$^{-1}$ in irradiated crystal implies the relative disordering of the molecules in the damaged volume along the ion tracks in single crystal [32]. The bands in high wavenumber (3000 – 3500 cm$^{-1}$) arise mainly due to N–H stretching vibrations and at around 500 cm$^{-1}$ corresponding to the $\delta_{as}$(N – C – N), they are also strongly affected by SHI.

### 3.2. Powder X-ray diffraction analysis

The X-ray diffraction patterns recorded for the face corresponding to (200) diffraction planes of pristine as well as irradiated single crystals are shown in Fig. 2 (a) and (b). In both the spectra only two peaks at 15.97 and 32.22 degree of 2-theta are observed which are corresponding to (200) and (400) diffraction planes. The slight broadening of peaks in the irradiated crystal indicates the damage to crystalline quality which is further elucidated in detail by HRXRD analysis in the following section.

### 3.3. High resolution X-ray diffraction analysis

The HRXRD rocking curves recorded for (200) diffraction planes of as grown and irradiated crystals are shown in Figs. 3 (a) and (b). The RC of pristine crystal has single peak and its FWHM (full width at half maximum) is 16 arcs (Fig. 3 (a)). The RC is almost symmetric with respect to the peak position and the diffracted intensity falls sharply on both the sides.

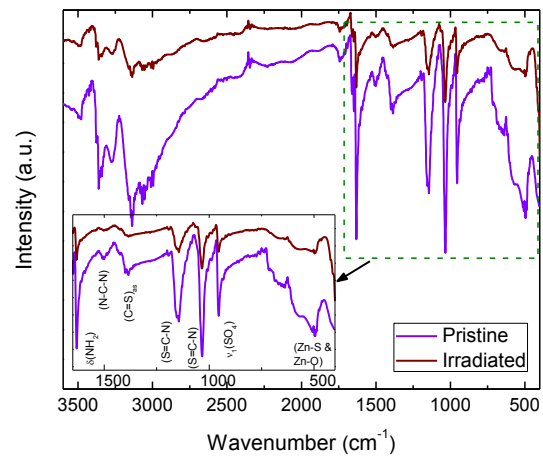

**Fig. 1.** The FTIR spectra of pristine (violet line) and irradiated (brown line) ZTS crystals recorded in reflection mode for (100) crystal face. Inset shows the magnified view of the bands in the range of 400-1650 cm$^{-1}$ along with assigned bonds and functional groups.

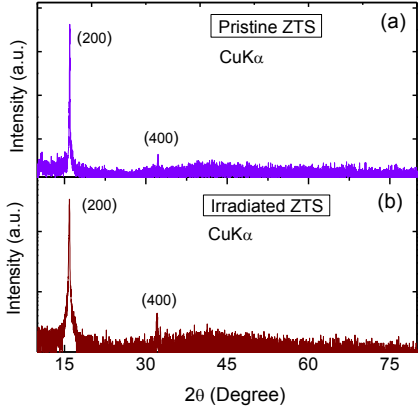

**Fig. 2.** The Powder X-ray diffraction spectra recorded for (100) face of (a) pristine and (b) irradiated single crystals.

These features indicate that the crystal is almost perfect and free from the structural defects such as grain boundaries and dislocations. However, the presence of nominal concentration of point defects and their agglomerates cannot be ruled out. The FWHM of the RC is close to that expected from the plane wave dynamical theory of X-ray diffraction for a perfect single crystal [33]. On close observation of RC, one can see that with respect to the peak position, the scattered intensity in the −ve side is slightly higher in comparison to that of +ve side. Such a slight asymmetry of RC depicts that the pristine crystal contains a few vacancy defects. This can be well understood from the fact that the lattice around the vacancy defect core undergoes the expansion and lead to the increase of lattice parameter $d$. This results in higher scattered intensity at lower angles with respect to the exact peak position according to Bragg's condition ($2d\sin\theta_B = n\lambda$, where $d$ is the lattice parameter and $\lambda$ is the wavelength of incident X-ray beam). The schematic of the lattice expansion around the vacancy defect core is given in the inset of Fig. 3 (a).

For better understanding of the crystalline perfection, the theoretical rocking curves corresponding to the (200) diffraction planes of ZTS are deduced and shown in Fig. 4. These RCs are deduced in symmetric Bragg reflection by assuming an ideally perfect crystal lattice of ZTS. These two RCs are known as Darwin and Darwin-Prince curves, the linear absorption correction for X-ray is taken into account for the later one. The FWHM ($\Delta\theta_{1/2}$) of such curves can be readily obtained from the following equation;

$$\Delta\theta_{1/2} = \frac{2\tau}{\sin 2\theta_B} F'_H |P|$$

(1)

here, $\tau = r_e\lambda^2/\pi V$, $r_e$ is the classical electron radius, $V$ is the volume of the unit cell, $\theta_B$ is the Bragg angle, $F'_H$ is the real component of the structure factor for (h k l) reflection and $P$ is the polarization factor which is taken as unity. The FWHM of these curves is 0.28 arc s which is very less, such a low FWHM and high reflectivity for ZTS is expected because of the light elements present in ZTS crystal, which is otherwise higher i.e. ~9.5 arc s for CdTe, a heavy element crystal.

The RC of irradiated crystal (Figs. 3 (b)) has quite different features compared to that of pristine crystal. In comparison to

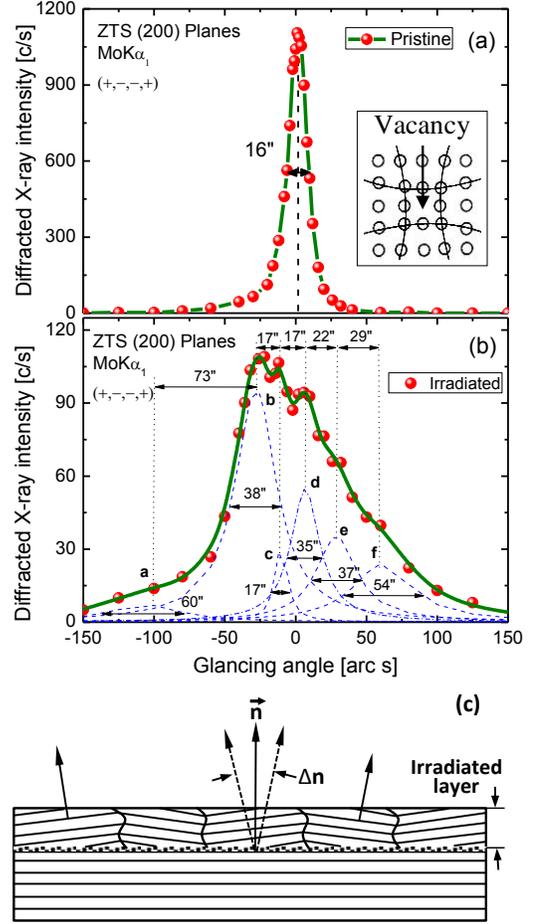

**Fig. 3.** The high resolution XRD rocking curves for the (200) diffraction planes of (a) pristine ZTS with an inset showing the vacancy defect depicting the expansion of lattice around the defect core, (b) rocking curve for (200) diffraction planes of irradiated ZTS, and (c) the schematic representation of irradiated crystal which contains a highly damaged top layer in which different parts have different normal with a distribution **Δn** around the normal **n** of main crystal.

RC of pristine, it has large FWHM of 84 arc s, with high scattered intensity along the tails up to long range of glancing angle on both sides of the exact peak position. Instead of a single peak it has multiple peaks; however, these peaks are not well-resolved from each other. This feature of RC depicts the formation of mosaic structures in the irradiated region of the crystal. The convoluted curve (solid green line) well fitted with the experimental points (solid orange circles) is obtained by Lorentzian fit. On deconvolution of this curve we get six peaks designated by letters as **a**, **b**, **c**, **d**, **e** and **f**. The peak c may be considered as main peak corresponding to the main crystal domain and its low FWHM (17 arc s) depicts that the main crystallite is still having good quality. The five additional peaks are corresponding to grain boundaries caused by SHI. The **a** and **b** are 17 and 90 arc s away from the main peak at lower angle side, and peaks **d**, **e** and **f** are 17, 39 and 68 arc s away from the main peak at higher angle side. The peaks **b**, **d** and **e** indicate very low angle grain boundaries, and the peaks **a** and **f** indicate low angle grain boundaries. It is important to mention here that the angles of misorientation of these boundaries are very low and in the convoluted curves these are not well resolved, which depicts that the crystal blocks corresponding to these peaks are in the form of mosaic blocks. It is interesting to observe here that FWHM of additional peaks corresponding to the grain boundaries are below 60 arc s which indicate that the mosaic blocks contain point defects and their agglomerates [34].

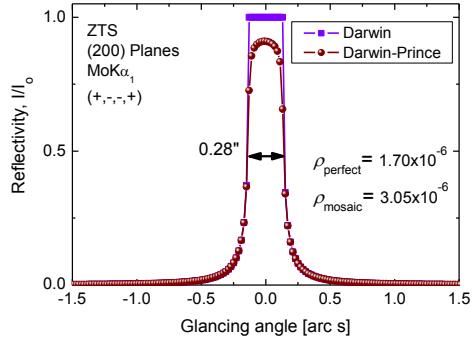

**Fig. 4.** The Darwin (violet solid squares and line) and Darwin-Prince (brown solid circles and line) theoretical rocking curves deduced for (200) diffraction planes of pristine ZTS crystal.

The Fig. 3 (c) is a schematic representation of irradiated ZTS crystal; the top layer shows the formation of mosaic blocks due to SHI irradiation. Beneath this, a deposition/implantation layer is shown by dotted region. For ZTS crystal lattice, the projected range of 150 MeV Au SHI has been estimated using SRIM (stopping range of ion in matter) software and found to be ~20$\mu$m. The pristine crystal, in the absence of structural grain boundary has only one diffraction vector for (200) diffraction planes, whereas, the irradiated crystal possess the mosaic blocks and have the diffraction vectors for (200) planes of these blocks which are not collinear but have some angular distribution due to the misorientation of these blocks. The broad angular range of RC depicts the misorientation of the mosaic blocks. The $\omega$-scan with highly monochromatic ($\Delta\lambda/\lambda \sim 10^{-5}$) and parallel beam (divergence << 3 arc s) for RCs provides the special advantage, as the local diffraction peaks from all the misoriented mosaic structures can be collected in one scan, which is otherwise not possible by $\omega$-$2\theta$ scan wherein diffraction from one block may come in one scan and the estimate of tilt angles from very low angle boundaries could not be possible. Formation of such structural mosaics/grain boundaries is possible because the SHI transfer huge amount of energy to the crystal lattice with heavy disturbance to the lattice and at suitable energies, the disturbed localized regions may anneal due to in-situ thermal energy generated during the process of implantation of heavy ions. In addition to the FWHM, the area under the RC also known as integrated intensity ($\rho$) gives an additional understanding about the degree of crystalline perfection of the crystal. For an ideally *imperfect* (or *mosaic*) crystal the *kinematical* theory is found to be suitable and with consideration of the linear absorption coefficient ($\mu$), $\rho_{mosaic}$ can be expressed as [35,36];

$$\rho_{mosaic} = \frac{N^2 \lambda^3}{2\mu} |F_H|^2 \left(\frac{e^2}{mc^2}\right)^2 \frac{1+|\cos 2\theta|^2}{2\sin 2\theta} \quad (2)$$

Here, $N$ and $F_H$ are number of unit cells and the structure factor, $e^2/mc^2$ is the classical electron radius = $2.818 \times 10^{-13}$ cm ($e$ and $m$ are the charge and mass of the electron and $c$ is the velocity of light).

For the large perfect single crystal the *dynamical* theory of X-ray diffraction [35,37] must be used, which takes into account the interactions between waves scattered from all irradiated atoms or scattering units. Any atom inside the crystal receives scattered waves from all the other atoms in addition to the incident beam (after attenuation due to scattering and absorption). Accordingly, $\rho_{perfect}$ for an ideally *prefect* crystal is given by [35,37]:

$$\rho_{perfect} = \frac{8}{3\pi} N\lambda^2 |F_H| \left(\frac{e^2}{mc^2}\right) \frac{1+|\cos 2\theta|}{2\sin 2\theta} \quad (3)$$

One important difference between Eqns (2) and (3) is that in case of ideally *mosaic* crystals, the integrated intensity is proportional to the square of $F_H$ whereas in ideally *perfect* crystals, it is simply proportional to the magnitude of $F_H$. Because of this clear contrast, the experimental value of $\rho_{mosaic}$ is expected to be much more than that of $\rho_{perfect}$. In case of ZTS for (200) diffraction planes, the theoretically calculated values of $\rho_{mosaic}$ and $\rho_{perfect}$ respectively are $3.05 \times 10^{-6}$ and $1.70 \times 10^{-6}$ rad. Such a behaviour of $\rho$, with quality of crystal, has been well discussed in our recent investigation on Czochralski grown Benzophenone single crystals, in which $\rho_{observed}$ was significantly higher for the crystal having mosaic blocks over to that of almost perfect crystal without any mosaic structure or grain boundaries [38]. However, in the present case, $\rho_{observed}$ for experimental RCs of pristine and irradiated ZTS crystals exhibits almost the same value. This anomaly in $\rho_{observed}$ may be the consequence of incoherent scattering at the heavily damaged top layer for the scattered intensity that is originated from the relatively more perfect lower regions, which leads to considerable absorption [39]. Due to scattering at the surface, there is a possible reduction of receiving intensity also from the relatively more perfect lower regions of the crystal. The depression in the diffracted intensity due to the absorption and incoherent scattering clearly depicts the formation of lattice defects induced on the surface by SHI. It is also interesting to observe a quite high scattered intensity over long angular range at positive side of the peak position of irradiated crystal, this feature signifies the presence of interstitial point defects and their agglomerates of good density [34], owing to the occupancy of SHI at the interstitial sites in crystal lattice. These defects result in the compression of lattice very close to the defect core and give rise to higher scattered intensity in positive side of RC [22,40].

*3.4. Raman analysis*

The ZTS is a non-centrosymmetric crystal and 360 vibrational modes are possible corresponding to the different atoms present in unit cell [41,30]. There are 3 acoustic modes and 357 optical modes; the acoustic modes are Raman active. The recorded Raman spectra for (200) face of pristine as well as irradiated crystals are shown in Fig. 5. Both the spectra were recorded in reflection mode under the identical experimental conditions at room temperature. Both the spectra look different, the pristine crystal has well defined peaks with very low scattered background intensity. The irradiated crystal has very high background which indicates the amorphization of top layer of crystal due to SHI irradiation [15,16]. All the observed peaks are consistent with the crystal symmetry and shown in Table I.

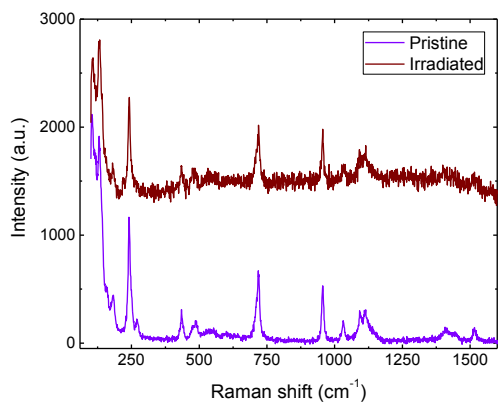

**Fig. 5.** The Raman spectra of pristine (violet solid line) and irradiated (brown solid line) crystals recorded for the (100) crystal faces.

**Table 1.** The observed Raman modes for the pristine and irradiated ZTS crystals

| Raman band (cm$^{-1}$) | | Assignments |
| --- | --- | --- |
| Pristine | Irradiated | |
| 103, 113, 117, 120 | 103, 113, 117, 120 | Lattice vibrations |
| 130, 132 | 130, 132, 140 | Lattice vibrations |
| 157 | 157 | Lattice vibrations |
| 182 | 180, 222 | Lattice vibrations |
| 240 | 241 | Lattice vibrations |
| 265, 270 | - | Zn–S and Lattice vibrations |
| 435 | 434 | $v_2(SO_4)$ |
| 476, 486 | 476, 486 | $\delta(N–C–N)$ |
| 532, 552 | 532, 552 | $\delta(NH_2)$, $v(NH_2)$ of Zn–S and Zn–O |
| 590 | - | $v(Zn–O)$ |
| 602 | - | $v_4(SO_4)$ |
| 646 | 646 | $v_4(SO_4)$ |
| 718 | 718 | $v_s(C–N)$ |
| 956 | 956 | $v_1(SO_4)$ |
| 1031 | 1031 | S=C–N |
| 1057 | 1057 | S=C–N |
| 1092 | 1092 | $\rho(NH_2)$ |
| 1114 | 1114 | $v_3(SO_4)$ |
| 1139 | 1039 | S=C–N |
| 1408 | 1401 | $v_{as}(CS)$ |
| 1444 | - | $v_{as}(CS)$ |
| 1516, 1522 | 1516 | S=C–N, $v_{as}(CN)$ |
| 1655 | 1648 | $\rho(NH_2)$ |

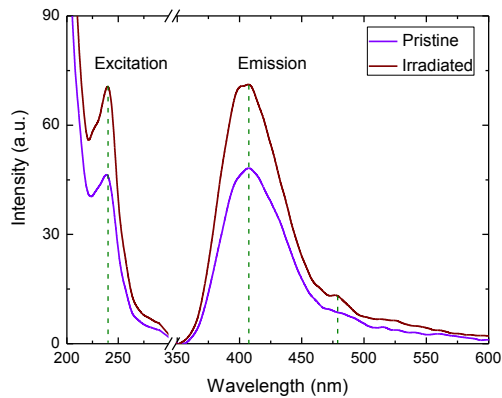

**Fig. 6.** The photoluminescence excitation and emission spectra recorded for (100) crystal face of pristine (violet solid line) and irradiated (brown solid line) ZTS crystals. The dotted lines show the position of excitation and emission peaks.

The peaks below 300 cm$^{-1}$ are corresponding to the lattice vibrations; peaks around 270 cm$^{-1}$ owing to Zn–S stretching in pristine ZTS are almost disappeared in irradiated crystal, which indicates the breaking and disordering of molecular bonds by SHI, hence bonding of thiourea with Zn is significantly influenced. The peaks above 1000 cm$^{-1}$ are broadened and their intensity is heavily reduced, indicating the damage and disordering of bonds in urea molecules and its coordination bonding in ZTS lattice. It is evident that SHI lead distortion of lattice vibrations, however, no other phase formation is observed. It is worth to mention here that Raman spectra were recorded in reflection mode, therefore, the above analysis may be restricted only to damaged top layer of crystal in irradiated region.

*3.5. Photoluminescence*

The PL excitation and emission spectra for pristine as well as irradiated crystal specimen are shown in Fig 6. The spectra were recorded for the crystal face of (200) diffraction planes. In the excitation spectra, the strong absorption has been observed at ~ 239 nm. The optical density for excitation spectra of irradiated crystal increased, however, no shift in the peak position observed. This considerable increase in the optical density for irradiated crystal gives the direct consequence of presence of structural defects in large number. For both the crystals strong emission peak at ~406 nm has been observed. The strong emission for pristine crystal in violet region is attributed to the intrinsic defect levels which develop during the growth process and are generally unavoidable [42,43]. For irradiated crystal the emission intensity increased significantly which depicts the formation of large number of luminescent color centres corresponding to the defects induced by SHI. The enhanced emission intensity for irradiated crystal is directly related with high optical density for excitation. In our earlier studies on LiF single crystal, the enhancement in PL intensity has been well correlated with point defects [44]. For irradiated crystal, a small additional peak at ~ 478 nm can be clearly seen, which may be attributed to the origination of additional spectroscopic states corresponding to the lattice defects induced by SHI. This is in agreement with the defect related PL emission observed in $Cr^{3+}$ doped ZTS single crystal which was well-correlated with the HRXRD investigations [45].

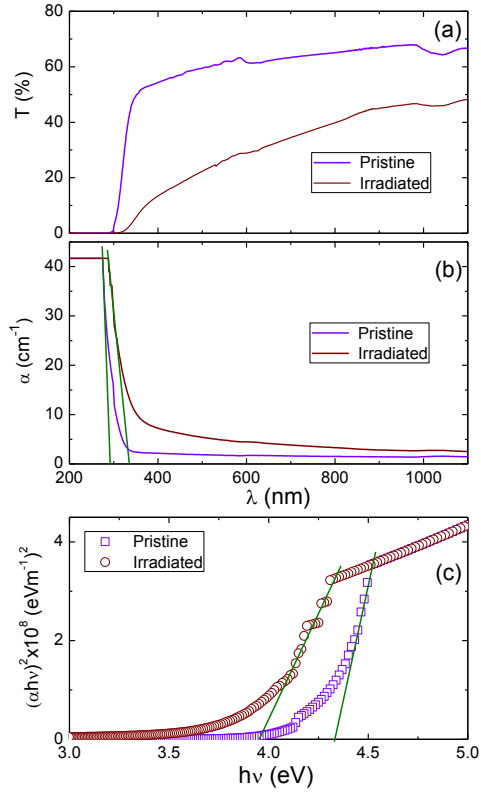

**Fig. 7.** (a) The optical transmission, (b) absorption coefficient and (c) $(\alpha h\nu)^2$ vs. $h\nu$ spectra for pristine (violet open squares) and irradiated (brown open circles) ZTS crystals.

*3.6.    UV-VIS-NIR analysis*

The UV-Vis-NIR transmittance spectra for the pristine and irradiated crystals are recorded in wavelength range of 200-1100 nm as shown in Fig. 7 (a). The pristine crystal is highly transparent over the entire wavelength range above 300 nm. The high transmittance of pristine crystal indicates its high crystalline quality and absence of any major defects like structural grain boundaries. For irradiated crystal, the transmittance reduced to a great extent in the entire region. This decrease in transparency is more prominent for the lower wavelengths, compared to that for higher wavelengths. This decrease in the transparency of irradiated crystal is attributed to the structural defects induced by the SHI as revealed by HRXRD. This is due to the fact that scattering of lower wavelength photons from the defects is high compared to those of higher wavelength photons. The SHI induced defects also act as the trapping centers for the photons. It is interesting to observe that in case of the irradiated crystal, the transmittance does not fall sharply but decreases slowly over a long wavelength range. Such decrease in optical transparency may be due to the fact that mosaics present in the irradiated region of the crystal lead to the multiple refraction and scattering of the incident beam at the boundaries [38]. A significant red-shift of ~35 nm in the cut-off wavelength has been observed. This red-shift and decrease in the transmittance may be attributed to the origination of localized energy states corresponding to the structural defects induced by SHI, in the band gap of ZTS crystal.

The absorption coefficient ($\alpha = A/l$; here, $A$ is absorbance and $l$ is the optical path-length or the thickness of crystal) was evaluated by recording the absorbance spectra and has been plotted with wavelength [Fig. 7 (b)]. The spectra show that the irradiated crystal has higher $\alpha$ over the entire investigated wavelength range, may be due to the fact that the defects trap and scatter the photons and results in high absorbance. However, this increase in $\alpha$ is more prominent for the lower wavelengths. The optical band gap ($E_g$) has been evaluated using the plot for $(\alpha h\nu)^2$ vs. $h\nu$ [Fig. 7 (c)] using the following relation:

$$(\alpha h\nu)^2 = A(E_g - h\nu)$$

(4)

Here, $E_g$ represents the optical band-gap and $A$ is a constant [46]. The values of $E_g$ have been evaluated by extrapolating the linear part of the plots to abscissa ($h\nu$) as shown in Fig. 7 (c). The values of $E_g$ for pristine and irradiated crystals are respectively 4.3 and 3.9 eV. The decrease in the optical band gap of irradiated crystal may be attributed to the formation of SHI induced defect related localised energy states in the energy-gap [47]. The localized states in the energy-gap generally occur for the crystals having extrinsic defects or disorders [48] and the defect related band gap variation has been discussed in our recent study on ZTS crystals [45]. The appearance of additional emission peak at ~478 nm for irradiated crystal also indicates the presence of defects related localized energy states.

**4. Conclusion**

In summary, we conclude that $Au^{9+}$ SHI irradiation resulted in the significant breaking and disordering of molecular bonds and functional groups of ZTS single crystal. From powder XRD analysis it is evidenced that no additional crystalline phase was formed due to irradiation. HRXRD investigations revealed: (i) the pristine ZTS crystal was almost perfect and free from the structural grain boundaries and dislocations, (ii) irradiation of crystal with 150 MeV Au SHI lead to the formation of mosaics or very low angle structural grain boundaries, and (iii) the mosaics contain the point defects and their agglomerates with good density. The irradiation significantly affected the lattice vibrations, molecular bonds and functional groups of ZTS, but do not change the crystal phase. It is found that SHI significantly enhance PL emission of ZTS in blue region of wavelengths which may be suitable for photonic applications where a selective high contrast blue emission is prerequisite. The SHI are capable to modify optical transparency and band gap and hence depicts that the irradiation of ZTS crystals may be useful to tailor the optical properties for optical window applications.

**Acknowledgments**

The authors thank the Director, NPL, India, for continuous encouragement in carrying out this work and the Director,

IUAC, New Delhi, for the access to perform irradiation experiments on 15-UD Pelletron. One of the authors, SKK, acknowledges CSIR, Ministry of Science and Technology, Government of India, for the grant of Senior Research Fellowship [31/1(293)/2008-EMR-I].